\documentclass[fleqn,twoside]{article}
\usepackage{espcrc2}

\usepackage{epsfig,psfig}
\input epsf

\newcommand{\AmS}{{\protect\the\textfont2
       A\kern-.1667em\lower.5ex\hbox{M}\kern-.125emS}}
\def\be{\begin{equation}}
\def\ee{\end{equation}}
\def\bea{\begin{eqnarray}}
\def\eea{\end{eqnarray}}
\def\beas{\begin{eqnarray*}}
\def\eeas{\end{eqnarray*}}
\newcommand{\D}{\displaystyle}

\title{Modeling Quasi-elastic Form Factors for Electron and Neutrino Scattering}

\author{H. Budd\address[roc]{Department of Physics and Astronomy,
             University of Rochester,
             Rochester, New York 14618,  USA}, A. Bodek\addressmark[roc]
	and 
	J. Arrington\address[arg]{Argonne National Laboratory, Argonne,
 Illinois 60439, USA}}
\begin{document}

\begin{abstract}
We calculate the total and differential
quasielastic cross sections for neutrino and antineutrino scattering on nucleons using
up to date fits to the nucleon elastic electromagnetic form factors 
$G_E^p$, $G_E^n$, $G_M^p$, $G_M^n$,
and weak and pseudoscalar form factors. We find that using the updated 
non-zero value of $G_E^n$ 
has a significant effect on both the total and differential neutrino and antineutrino quasielastic
cross sections. Previous extractions of the weak axial form factor from neutrino scattering data are
sensitive to the assumptions that were used for the vector form factors. We perform a re-analysis of 
previous neutrino data using updated form factors and obtain updated value of the axial vector mass.
 (Presented by Howard Budd at NuInt02,   Dec. 2002, Irvine, CA, USA~\cite{nuint02})
\vspace{1pc}
\end{abstract}

\maketitle

\section{INTRODUCTION}

Experimental  evidence for oscillations among the three
neutrino generations has been recently reported~\cite{Fukada_98}.
Since quasielastic (QE) scattering forms an important component of 
neutrino scattering
at low energies, we have undertaken to investigate 
QE neutrino scattering using the latest information
on nucleon form factors. 

Recent experiments at SLAC and Jefferson Lab (JLab) have 
given precise measurements of the vector electromagnetic
form factors for the proton and neutron. 
These form factors can be related to the form factors for QE
neutrino scattering by conserved vector current hypothesis, CVC. 
These more recent form factors can be used to give better predictions
for QE neutrino scattering.

\section{EQUATIONS FOR QE SCATTERING}

The hadronic current for QE neutrino scattering is given by~\cite{Lle_72}
\begin{eqnarray*}
 \lefteqn{<p(p_2)|J_{\lambda}^+|n(p_1)>  =   } \nonumber \\
& \overline{u}(p_2)\left[
  \gamma_{\lambda}F_V^1(q^2)
  +\frac{\D i\sigma_{\lambda\nu}q^{\nu}{\xi}F_V^2(q^2)}{\D 2M} \right. \nonumber \\
& \left. ~~~~~~~~~~~+\gamma_{\lambda}\gamma_5F_A(q^2)
+\frac{\D q_{\lambda}\gamma_5F_P(q^2)}{\D M} \right]u(p_1),
\end{eqnarray*}
where $q=k_{\nu}-k_{\mu}$, $\xi=(\mu_p-1)-\mu_n$, and 
$M=(m_p+m_n)/2$.  Here, $\mu_p$ and $\mu_n$ are the 
proton and neutron magnetic moments.
We assume that there are no second class currents, so the scalar
form factor  $F_V^3$ and the tensor form factor $F_A^3$
need not be included. 
Using the above current, the cross section is
\begin{eqnarray*}
 \lefteqn{ \frac{d\sigma^{\nu,~\overline{\nu}}}{dq^2} = 
  \frac{M^2G_F^2cos^2\theta_c}{8{\pi}E^2_{\nu}}\times }  \nonumber \\
&\left[A(q^2) \mp \frac{\D (s-u)B(q^2)}{\D M^2} + \frac{\D C(q^2)(s-u)^2}{\D M^4}\right],
\end{eqnarray*}
where
%$$ s-u = 4ME_{\nu} + q^2 - m_l^2$$
\begin{eqnarray*}
\lefteqn{A(q^2)= 
 \frac{m^2-q^2}{4M^2}\left[
    \left(4-\frac{\D q^2}{\D M^2}\right)|F_A|^2 \right.} \nonumber \\
& \left.  -\left(4+\frac{\D q^2}{\D M^2}\right)|F_V^1|^2
  -\frac{\D q^2}{\D M^2}|{\xi}F_V^2|^2\left(1+\frac{\D q^2}{\D 4M^2}\right) \right.  \nonumber \\ 
& \left.  -\frac{\D 4q^2ReF_V^{1*}{\xi}F_V^2}{\D M^2} 
    \right],
%\right.  \\    &   & \left.
%  -\frac{m_l^2}{M^2}\left(|F_V^1+{\xi}F_V^2|^2
%    \right)
\end{eqnarray*}
$$
B(q^2) = -\frac{q^2}{M^2}ReF_A^*(F_V^1+{\xi}F_V^2),
$$
$$
C(q^2) = \frac{1}{4}\left(|F_A|^2 + |F_V^1|^2 -
 \frac{q^2}{M^2}\left|\frac{{\xi}F_V^2}{2}\right|^2\right).
$$
Although we have
have not shown terms of order $(m_l/M)^2$, and 
terms including $F_P(q^2)$ (which
is multiplied by  $(m_l/M)^2$), these terms
%$F_P(q^2)$ and  terms of order  $(m_l/M)^2$ 
are included in our calculations~\cite{Lle_72}.)
The form factors $ F^1_V(q^2)$ and  ${\xi}F^2_V(q^2)$
are given by:
$$ F^1_V(q^2)=
\frac{G_E^V(q^2)-\frac{\D q^2}{\D 4M^2}G_M^V(q^2)}{1-\frac{\D q^2}{\D 4M^2}},
$$
$$
{\xi}F^2_V(q^2) =\frac{G_M^V(q^2)-G_E^V(q^2)}{1-\frac{\D q^2}{\D 4M^2}}.
$$

We use the CVC to determine $ G_E^V(q^2)$ and $ G_M^V(q^2)$ 
from  the electron scattering form factors
$G_E^p(q^2)$, $G_E^n(q^2)$, $G_M^p(q^2)$, and $G_M^n(q^2)$:

$$ 
G_E^V(q^2)=G_E^p(q^2)-G_E^n(q^2), 
$$
$$
G_M^V(q^2)=G_M^p(q^2)-G_M^n(q^2). 
$$

Previously, many neutrino experiment have assumed
that the vector 
form factors are 
described by the dipole approximation.
$$ G_D(q^2)=\frac{1}{\left(1-\frac{\D q^2}{\D M_V^2}\right)^2 },~~M_V^2=0.71~GeV^2$$
$$ 
G_E^p=G_D(q^2),~~~G_E^n=0,
$$ 
$$ 
G_M^p={\mu_p}G_D(q^2),~~~ G_M^n={\mu_n}G_D(q^2).
$$
%table 1
\begin{table}
\begin{center}
\begin{tabular}{|l|c|}
\noalign{\vspace{-8pt}} \hline
$g_A$ & -1.267    \\
$G_F$ & 1.1803${\times}10^{-5}$ GeV$^{-2}$  \\
$\cos{\theta_c}$ & 0.9740 \\
$\mu_p$ & 2.793 $\mu_N$ \\
$\mu_n$ & -1.913 $\mu_N$ \\
$\xi$  & 3.706 $\mu_N$ \\
$M_V^2$ & 0.71 GeV$^2$\\ \hline
\end{tabular}
\end{center}
\caption{  The most recent values of the
parameters used in our calculations (Unless stated otherwise).}
\label{parameters}
\end{table}
We refer to the above combination of form factors
as `Dipole Form Factors'. It is an approximation that 
is improved upon in this paper. We will refer
to our updated form factors as `BBA-2003 Form Factors' 
(Budd, Bodek, Arrington).
Table~\ref{parameters} summarizes the
most up to date values of the coupling constants and magnetic
moments that we use in our calculations. Note that
$G_E^p$, $G_M^p$, and $G_E^n$ are positive, while
$G_M^n$ and the axial form factor $F_A$ are negative.

The axial form factor is given by 
$$ F_A(q^2)=\frac{g_A}{\left(1-\frac{\D q^2}{\D M_A^2}\right)^2 }. $$
This form factor needs to be extracted from 
QE neutrino scattering. However, at low $Q^2$,
this form factor can also be extracted from pion
electroproduction data. 

%figure1
\begin{figure}%[htb]
\begin{center}
\epsfxsize=2.91in
\mbox{\epsffile[65 470 550 710]{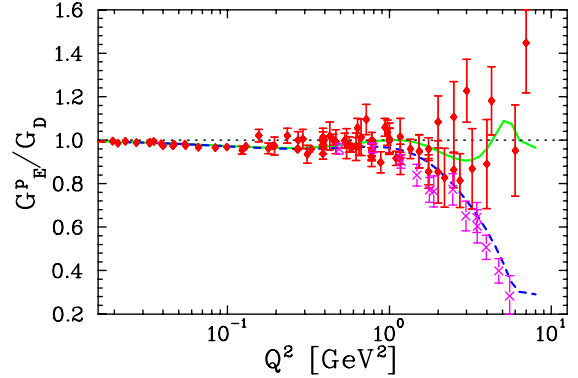}}
\end{center}
\caption{Our fits to $G_E^p/G_D$, using cross section data only
(solid), and with both the cross section and polarization transfer data
(dashed). The diamonds are the from Rosenbluth extractions
and the crosses are the Hall A polarization transfer data.  Note that
we fit to cross sections, rather than fitting directly to
the extracted values of $G_E^p$ shown here.}
\label{show_gep}
\end{figure}

Previous neutrino
experiments used $g_A$=$-1.23$, while
the best current value is $-1.267$. 
The world average from neutrino experiments for $M_A$  is
1.026 $\pm$ 0.020 GeV~\cite{Bernard_01}. 
The value of $M_A$ extracted
from neutrino experiments depends on both
the value of $g_A$ and the values of
the electromagnetic form factors 
which are assumed in the extraction process.
Since we are updating these form factors, 
new values of $M_A$ are
extracted from previous neutrino data
using the better known values for $g_A$ and the vector
form factors.

$M_A$ can also 
be determined from pion electroproduction,
which yields a world average value
of 1.069 $\pm$ 0.016 GeV~\cite{Bernard_01}. 
This value should
be reduced by 0.055 GeV when compared to  $M_A$ 
as measured in
neutrino data because of additional
corrections~\cite{Bernard_01}. Therefore,
pion electroproduction experiments predict
that $M_A$ should be 1.014 $\pm$ 0.016 GeV in
neutrino scattering.

In this communication, we
show that the value of 1.026 $\pm$ 0.02 GeV~\cite{Bernard_01}
as measured from the average of all neutrino scattering should
also be reduced by 0.025 GeV to account for incorrect vector
form factors used in the past. This corrected value of
1.001 $\pm$ 0.020 GeV is in good agreement with the 
theoretically corrected value 
from pion electroproduction of 1.014 $\pm$ 0.016 GeV.

From PCAC, the pseudoscalar form factor $F_P$ is
predicted to be
$$ F_P(q^2)=\frac{2M^2F_A(q^2)}{M_{\pi}^2-q^2}. $$
In the expression for the cross section,
$F_P(q^2)$ is multiplied by  $(m_l/M)^2$. 
Therefore, 
in muon neutrino interactions, this effect 
is very small except at very low energy, below 0.2~GeV.
The effect is larger, about $5\%$, for tau neutrino interactions.

%table 2
\begin{table*} 
\begin{center}
\begin{tabular}{|l|c|c|c|c|c|c|c|}
\noalign{\vspace{-8pt}} \hline
        & data     & $a_2$ & $a_4$ & $a_6$   & $a_8$     & $a_{10}$  &  $a_{12}$     \\ \hline
$G_E^p$ & CS + Pol & 3.253 & 1.422 & 0.08582 & 0.3318    & -0.09371  & 0.01076      \\
$G_M^p$ & CS + Pol & 3.104 & 1.428 & 0.1112  & -0.006981 & 0.0003705 & -0.7063E-05  \\
$G_M^n$ &          & 3.043 & 0.8548 & 0.6806 & -0.1287   & 0.008912  &              \\ \hline
$G_E^p$ & CS       & 3.226 & 1.508 & -0.3773 & 0.6109    & -0.1853   & 0.01596      \\
$G_M^p$ & CS       & 3.188 & 1.354 & 0.1511  & -0.01135  & 0.0005330 & -0.9005E-05  \\ \hline
\end{tabular}
\end{center}
\caption{ The coefficients of the inverse polynomial fits for the 
$G_E^p$, $G_M^p$, and $G_M^n$. Fits using cross section data only, and
using both cross section data and the Hall A polarization transfer
data are shown separately.
Note that these different polynomials replace $G_D$ 
in the expression for $G_E^p$, $G_M^p$, and $G_M^n$. 
The first three rows of the table  along with the fit of 
$G_M^p$ Krutov {\em et. al. }~\cite{Krutov_02} 
(see text) will be referred to as `BBA-2003 Form Factors'.}
\label{JRA_coef}
\end{table*}

%figure2
\begin{figure}%[htb]
\begin{center}
\epsfxsize2.91in
\mbox{\epsffile[65 470 550 710]{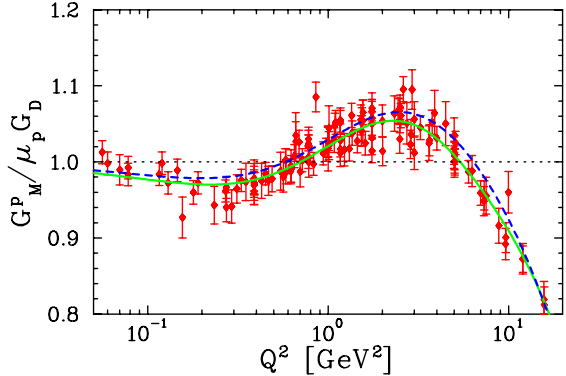}}
\end{center}
\caption{Our fits to $G_M^p/{\mu}_{p}G_D$.
The lines and symbols have the same meaning as Figure~\ref{show_gep}. }
\label{show_gmp}
\end{figure}

%figure3
\begin{figure}%[htb]
\begin{center}
\epsfxsize=2.71in
%\mbox{\epsffile[79  408  549  720]{show_gmn_ratio.eps}}
\mbox{\epsffile[65 470 550 710]{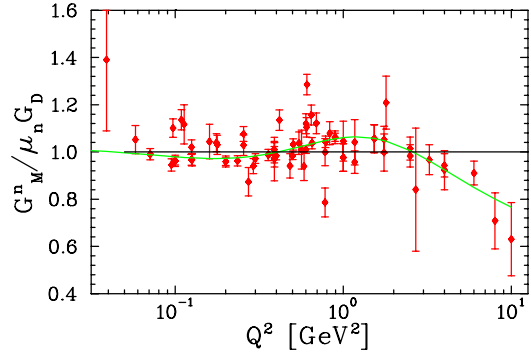}}
\end{center}
\caption{Our fit to $G_M^n/{\mu}_{n}G_D$.
The lines and symbols have the same meaning as Figure~\ref{show_gep}. }
\label{show_gmn}
\end{figure}

\section{UPDATED FORM FACTORS}

We have used an updated fit to the proton electromagnetic
form factors.  The fit is similar to the one described in
Ref.~\cite{JRA_03}, but using a slightly different fitting
function (described below), and including additional data
to constraint the fit at low $Q^2$ values.
Form factors can be determined from cross sections using
the standard Rosenbluth separation technique~\cite{JRA_03},
which is sensitive to radiative corrections, or from polarization
measurements using the newer polarization transfer technique~\cite{halla}.
The polarization 
measurements do not directly measure the form factors,
but measure the ratio $G_E$/$G_M$. 
Figures~\ref{show_gep},~\ref{show_gmp}, 
and~\ref{show_gmn} show the ratio of 
our fits divided by the dipole form, $G_D$.

\begin{figure}%[htb]
\begin{center}
\epsfxsize=2.91in
%\mbox{\epsffile[79  408  549  720]{show_gepgmp_ratio.eps}}
\mbox{\epsffile[65 470 550 710]{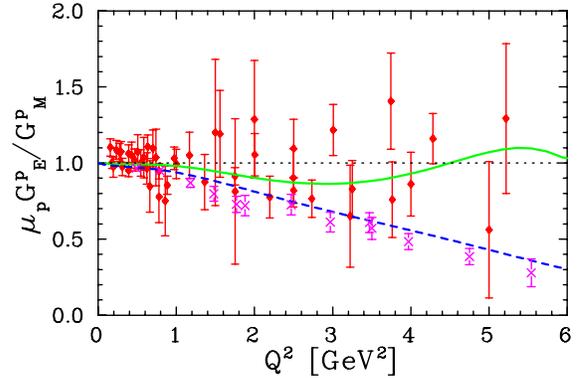}}
\end{center}
\caption{ Ratio of $G_E^p$ to $G_M^p$ as extracted by Rosenbluth
measurements and from polarization measurements.
The lines and symbols have the same meaning as Figure~\ref{show_gep}.  }
\label{show_gepgmp}
\end{figure}

Figure~\ref{show_gepgmp} shows our fits to $\mu_p G_E^p$/$G_M^p$.
The fit including only cross section data 
is roughly flat versus $Q^2$ ($Q^2=-q^2$), while ratio decreases with
$Q^2$ in the combined fit to
cross section and polarization transfer data. 
Although the polarization transfer measurement is
believed to have smaller systematic error, especially 
at high $Q^2$, the origin of this disagreement is not known.
If this disagreement comes from radiative corrections
to the electron, in particular two-photon
exchange terms, then the polarization transfer extraction
will give the correct ratio, but the overall scale
of $G_E^p$ at low $Q^2$ would be shifted down by
$\approx$3\%. Because the fit is constrained as $Q^2 \rightarrow 0$,
there will not be an overall shift in $G_E^p$ at low $Q^2$, but
there will be some uncertainty in the low $Q^2$ behavior.
Current experiments at JLab aim to better understand the source of the
disagreement by looking at the recoil proton in elastic electron-proton
scattering, thus minimizing the sensitivity to the dominant sources
of uncertainty in previous Rosenbluth separations.
However, since this discrepancy is most prominent at high $Q^2$, and
the fit is constrained at low $Q^2$, it has only
a relatively small effect on the neutrino QE scattering cross section.

To account for the fact that deviations from the dipole
form are different for each of the different form
factors, we fit  electron scattering data for each of the
form factors to an inverse polynomial
$$ G_{E,M}^{N}(Q^2)=\frac{G_{E,M}^{N}(Q^2=0)}{1+a_2Q^2+a_4Q^4+a_6Q^6+...}. $$
Table~\ref{JRA_coef} shows the parameters of our fit. 
We have done fits using cross section data
only (for the proton) and fits using both  both cross section data 
and polarization transfer data from JLab Hall A. For $G_E^p$,
the parameters in Table~\ref{JRA_coef} are used for $Q^2 < 6$~GeV$^2$.
For $Q^2>6$~GeV$^2$, the ratio of $G_E^p/G_M^p$ is assumed to be constant:
$$ G_E^p(Q^2) = G_M^p(Q^2) \frac{G_E^p(6~\mbox{GeV}^2)}{G_M^p(6~\mbox{GeV}^2)} $$

%figure4
%figure 5
\begin{figure}%[htb]
\begin{center}
\epsfxsize=2.71in
%\mbox{\epsffile[79  408  549  720]{show_gen_new.eps}}
\mbox{\epsffile[65 470 550 710]{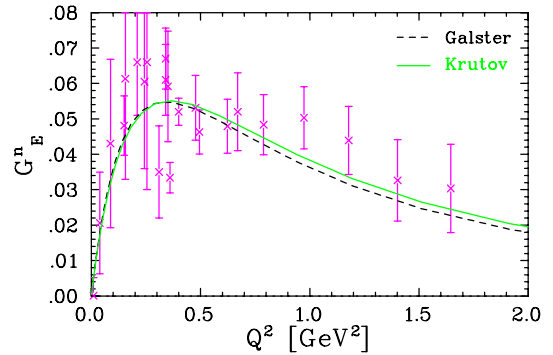}}
\end{center}
\caption{
Data and fits to $G_E^n$.  The dashed line is the Galster {\em et al.}
fit~\cite{Glaster_71}, and the solid line is the Krutov{\em et al.}
fit~\cite{Krutov_02}.}
\label{show_gen_new}
\end{figure}

Since the neutron has no charge, $G_E^n$ must be
zero at $q^2$=0, and previous neutrino experiments
assumed  $G_E^n(q^2)$=0 for all $q^2$ values.
However, it is non-zero away from $q^2$=0,
and its slope at $q^2$=0
is known precisely from neutron-electron scattering.
At intermediate $Q^2$, recent polarization transfer data give
precise values of $G_E^n(q^2)$.
Our analysis uses the parameterization of 
Krutov {\em et. al.}~\cite{Krutov_02}:
$$G_E^n(Q^2) = -\mu_n\frac{a\tau}{1+b\tau}G_D(Q^2),~~~\tau=\frac{Q^2}{4M^2},$$
with $a=0.942$ and $b=4.61$.
This parameterization is very similar to that of
Galster {\em et al.}~\cite{Glaster_71}, as shown
in Figure~\ref{show_gen_new}.

The first three rows of Table~\ref{JRA_coef},
along with the fit of 
$G_E^n$ of Krutov {\em et. al. }~\cite{Krutov_02}, 
will be referred to as `BBA-2003 Form Factors'.
For BBA-2003 Form Factors, both the cross 
section and polarization data are used in the extraction
of $G_E^p$ and $G_M^p$.

%figure6
\begin{figure}
\begin{center}
\epsfxsize=2.91in
\mbox{\epsffile[7 75 550 360]{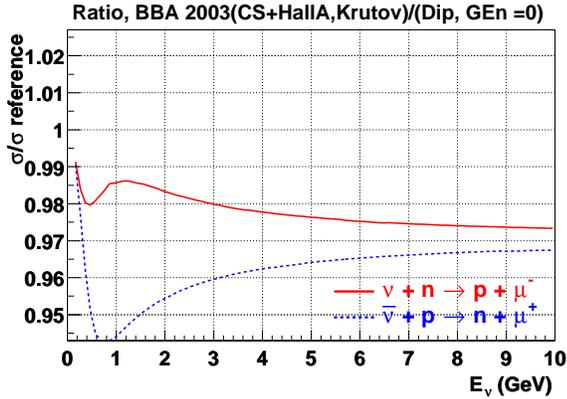}}
\end{center}
\caption{Ratio versus energy  of predicted neutrino (antineutrino)
QE cross section using BBA-2003 Form
Factors to the prediction using 
the dipole approximation with $G_E^n$=0.}
\label{ratio_JhaKJhaJ_D0DD}
\end{figure}

%%figure7
%\begin{figure}
%\begin{center}
%\epsfxsize=2.91in
%\mbox{\epsffile[7 75 550 360]{ratio_JKJJ_D0DD.eps}}
%\end{center}
%\caption{Ratio versus energy  of predicted neutrino (antineutrino)
% quasielastic cross section using updated form factors
%factors to the prediction using 
%the dipole approximation with $G_E^n$=0. The proton form
%factors are from the fit including only cross section data,
%while the neutron form factors are identical to those
%in the BBA-2003 Form Factors.}
%\label{ratio_JKJJ_D0DD}
%\end{figure}

\section{CROSS SECTIONS AND FITS TO $M_A$}
Figure~\ref{ratio_JhaKJhaJ_D0DD} 
shows the ratio versus neutrino energy of the
predicted neutrino (antineutrino) QE cross section
using our BBA-2003 Form Factors to the prediction using the
Dipole Form Factors. 
%
%Figure~\ref{ratio_JKJJ_D0DD} shows the ratio 
%vs neutrino energy of the
%predicted neutrino (antineutrino) QE cross section
%using our our most recent form factors {\it without
%polarization data} to the prediction using the
%Dipole Form Factors. The form factors for Figure~\ref{ratio_JKJJ_D0DD}
%use the same $G_M^n$ and $G_E^n$ as BBA-2003 Form Factors but only uses 
%only cross section data for $G_M^p$ and $G_E^p$.
%As seen from these figures
%the predicted cross sections (for neutrino
%and antineutrino QE) 
%scattering are very close if one either includes or does not include
%the Hall A polarization transfer data in the fits.
%
The same comparison was performed using the proton form factor extractions
that included
only the cross section data ({\it i.e.} excluding the polarization
results).  The results were nearly identical: the maximum difference between the cross
sections is less then 0.3\%.
This is because the form factors extracted
from the polarization transfer
data and from the electron scattering cross section data are different
only at high $Q^2$, while the neutrino cross sections are mostly
sensitive to  the form factors at low $Q^2$.

%figure7
\begin{figure}
\begin{center}
\epsfxsize=2.91in
\mbox{\epsffile[7 75 550 360]{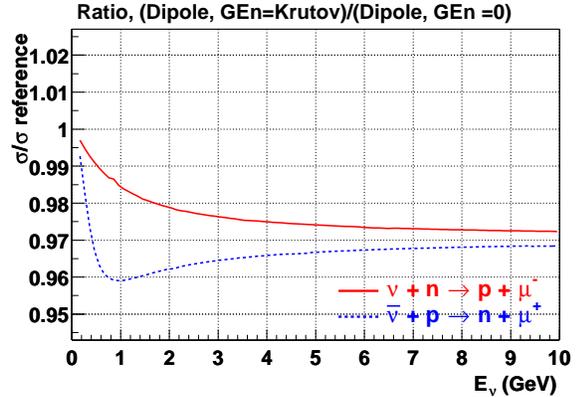}}
\end{center}
\caption{Ratio versus energy of the predicted
neutrino (antineutrino) QE cross section using 
$G_E^n$ from Krutov~\cite{Krutov_02} to the prediction using $G_E^n$=0.
In both cases, the dipole approximation
for the other form factors.}
\label{ratio_DKDD_D0DD}
\end{figure}
%figure8
\begin{figure}
\begin{center}
\epsfxsize=2.91in
\mbox{\epsffile[7 75 550 360]{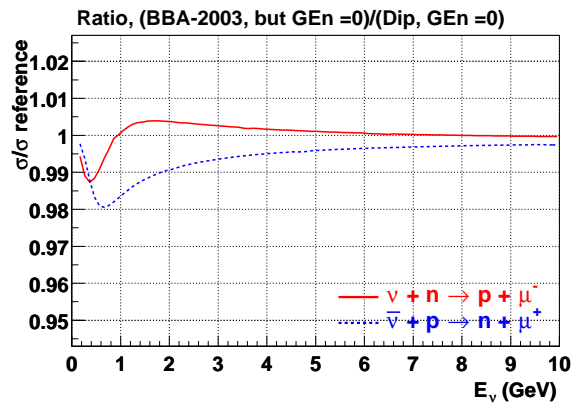}}
\end{center}
\caption{Ratio versus energy of 
predicted neutrino (antineutrino) QE cross section using 
our BBA-2003 Form Factor fits for $G_E^p$, $G_M^p$, and $G_N^n$ to the prediction
using the dipole approximation. In both cases, we use $G_E^n$=0.}
\label{ratio_Jha0JhaJ_D0DD}
\end{figure}
There is a large difference in the predicted neutrino (antineutrino)
cross section between using the BBA-2003 Form Factors and using
the Dipole Form Factor approximation.
The difference is 3\% at high energy and can become
as much as 6\% at 1 GeV.
As seen in 
Figure~\ref{ratio_DKDD_D0DD}, which shows the difference
between $G_E^n$ = Krutov and $G_E^n$ = 0, all the difference
at high energy and most of the the difference at low energy is due
to $G_E^n$. At the low energy region which is of interest
for neutrino  oscillation experiments, both a non-zero $G_E^n$ and
the deviations from the dipole form are important.  This is also the case
for the extraction of the axial form factor from neutrino data, since most
of the neutrino differential cross section data are at low $Q^2$.

  A 1\% increase in either $M_A$ or $|g_A|$ increases the
cross section about 1\%.  Replacing the old value of $g_A$=$-1.23$
with the more precise value of $g_A$=$-1.267$ increases the cross section 
by about 2.5\%.  Using the more recent value of $M_A$ of 1.02 instead of
the older value of 1.032 decreases the predicted  cross section about 1\%.
$F_P$ has almost no effect on the cross section except at very
low $E_{\nu}$. Therefore, even a very conservative error~\cite{Bernard_01}
on $F_P$  of $50\%$ has very little effect.

Previous neutrino measurements, mostly bubble chamber experiments,
extracted
$M_A$ using the best known assumptions at the time. Changing
these assumptions changes
the extracted value of $M_A$. Hence, $M_A$ needs to 
be updated using new form factors and up-to-date couplings. In this 
communication we will attempt to update the  
results from three previous  deuterium bubble  
chamber experiments. These are
Baker {\em et al.}~\cite{Baker_81}, Barish {\em et al.}~\cite{Barish_77}, 
Miller {\em et al.}~\cite{Miller_82}, 
and Kitagaki {\em et al.}~\cite{Kitagaki_83}. 
Barish {\em et al.} and  Miller {\em et al.}  are the 
same experiment, with the analysis of Miller {\em et al.} including
the full data set, roughly three times the statistics included in the
original analysis.
%Therefore, the calculation of the predicted cross sections
%by  Miller {\em et al.} 
%and Barish {\em et al.} should agree since Miller {\em et al.} is the 
%same experiment as Barish {\em et al.}

%figure9
\begin{figure}
\begin{center}
\epsfxsize=2.91in
\mbox{\epsffile[95 260 540 580]{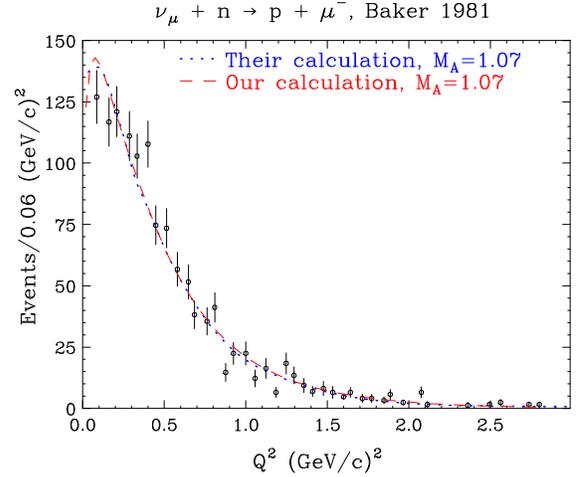}}
\end{center}
\caption{ $Q^2$ distribution from Baker {\em et al.}~\cite{Baker_81}.
The dotted curve is their calculation taken from their $Q^2$ distribution 
histogram. The dashed curve is our calculation using their assumptions. }
\label{Baker_81_ma107_nor100}
\end{figure}
%figure10
\begin{figure}
\begin{center}
\epsfxsize=2.91in
\mbox{\epsffile[95 260 540 580]{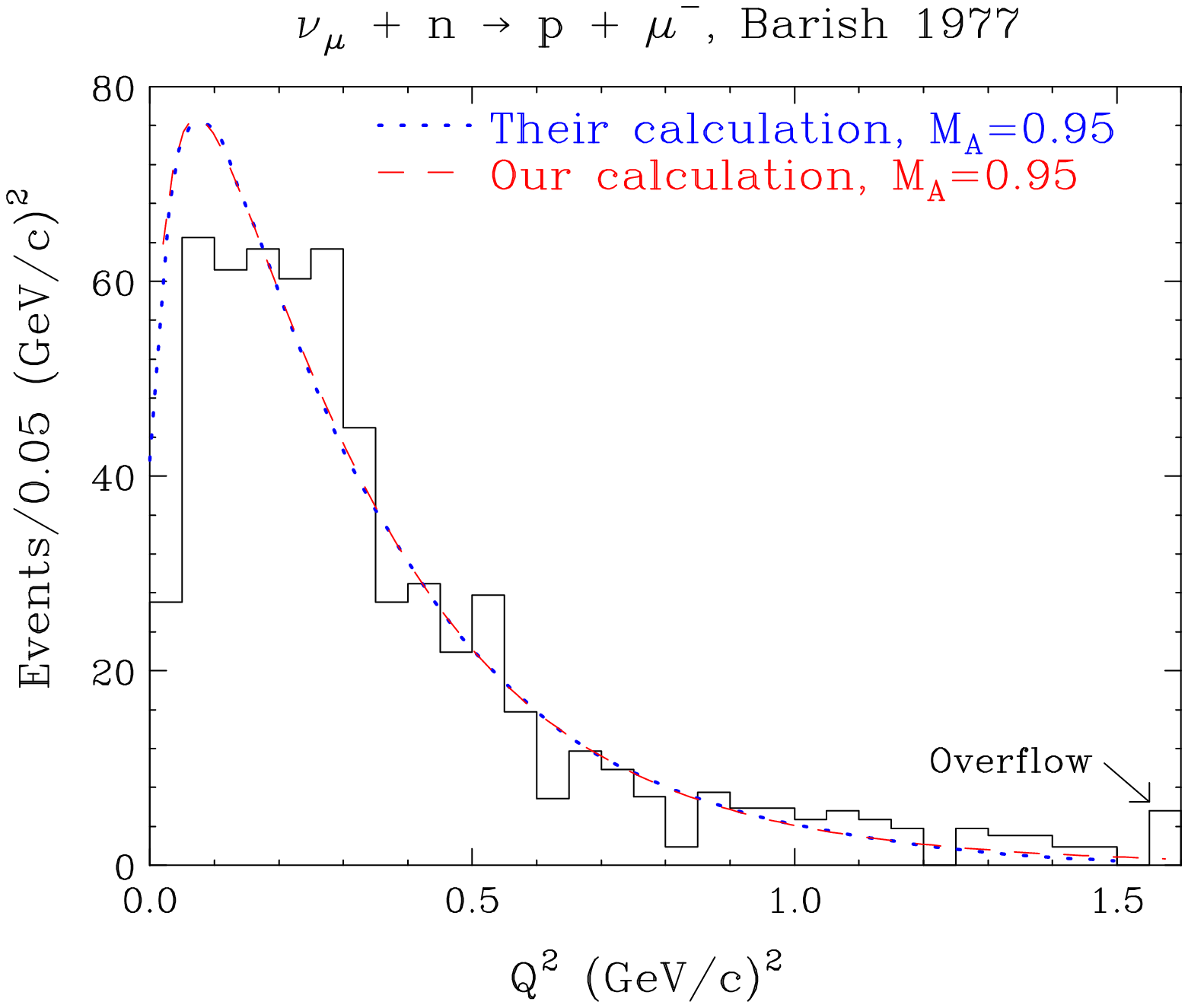}}
\end{center}
\caption{ $Q^2$ distribution from Barish  {\em et al.}~\cite{Barish_77}. 
The dotted curve is their calculation taken from their $Q^2$ distribution
histogram. The dashed curve is our calculation using their assumptions.}
\label{Barish_77_ma95_nor100}
\end{figure}

%figure11
\begin{figure}
\begin{center}
\epsfxsize=2.91in
\mbox{\epsffile[95 260 540 580]{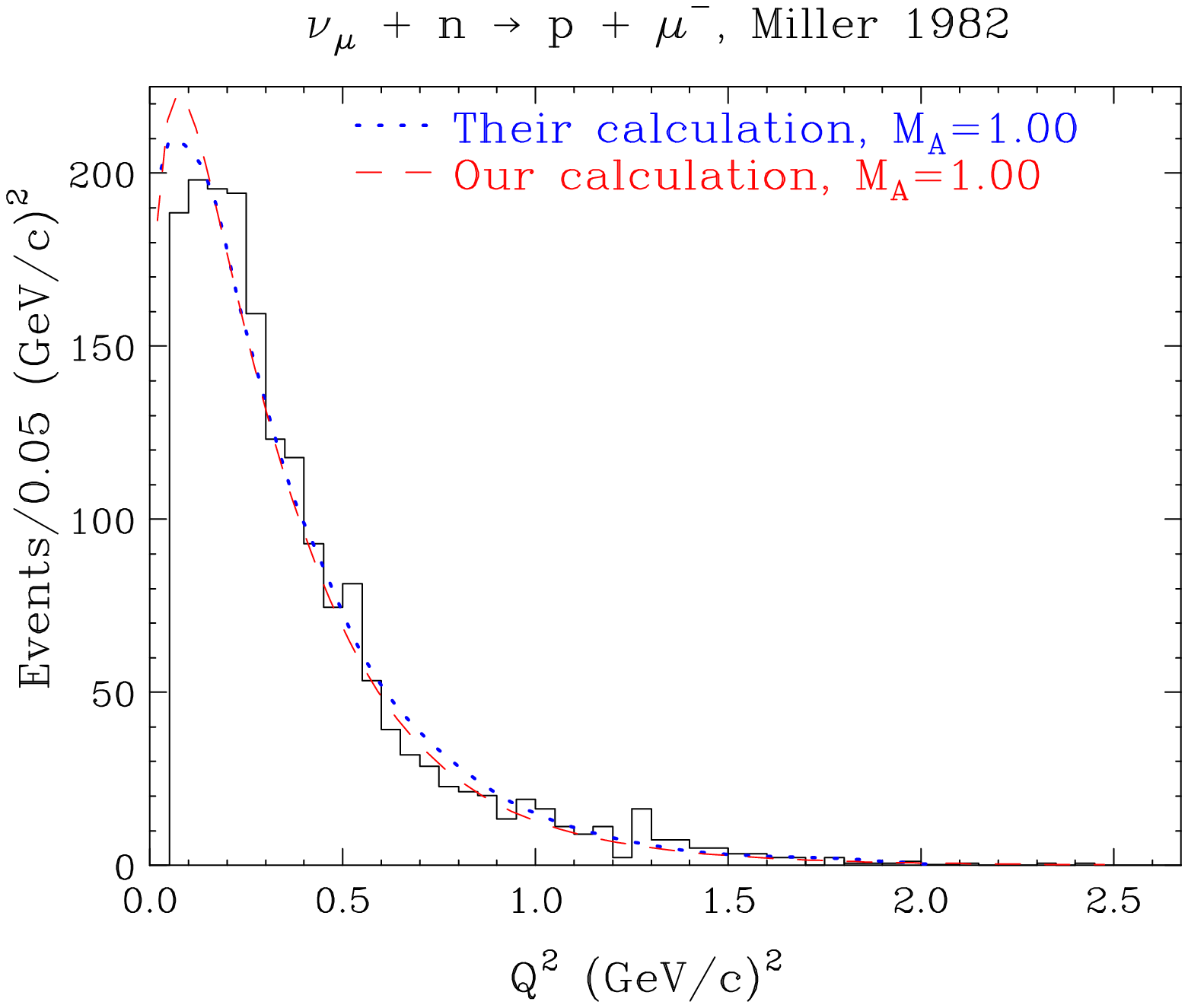}}
\end{center}
\caption{ $Q^2$ distribution from Miller {\em et al.}~\cite{Miller_82}.
The dotted curve is their calculation taken from their $Q^2$ distribution
histogram. The dashed curve is our calculation using their assumptions. }
\label{Miller_83_ma100_nor97}
\end{figure}
%figure12
\begin{figure}
\begin{center}
\epsfxsize=2.91in
\mbox{\epsffile[95 260 540 580]{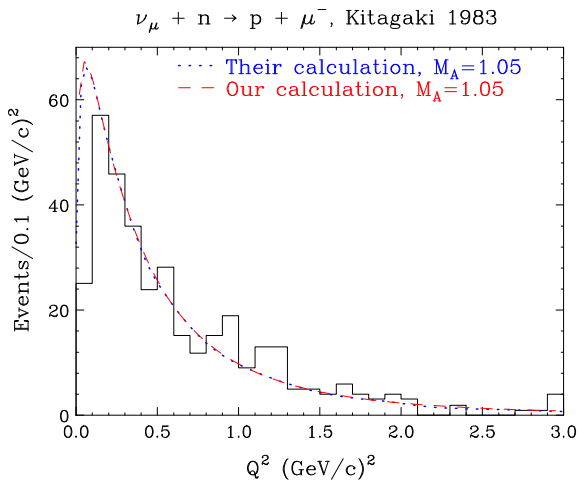}}
\end{center}
\caption{ $Q^2$ distribution from Kitagaki {\em et al.}~\cite{Kitagaki_83}.
The dotted curve is their calculation taken from their $Q^2$ distribution
histogram. The dashed curve is our calculation using their assumptions.  }
\label{Kit_83_ma105_nor95}
\end{figure}

We start by calculating the shape of the $Q^2$ distribution using the 
same form factors and couplings as used in the original extractions (including
$M_A$).
The flux is extracted from the flux figures shown in the original
papers, which we parameterize using a spline fit.
We extract the data and curves from
their publications by picking points off the plots, and 
fitting the points on the curves to a spline fit. 
These experiments did not use a pure dipole approximation, but
included a correction to the dipole form as parameterized by
Olsson {\em et al.}~\cite{Olsson_78}. 
They use $g_A$=$-1.23$, $M_V$=0.84 GeV (yielding $M_V^2$=0.7056 
GeV$^2$ instead of $M_V^2$=0.71 GeV$^2$), and 
a $D_2$ Pauli-suppression correction from Singh {\em et al.}~\cite{Singh_72}.
Our calculations and their calculations are
compared in Figures~ \ref{Baker_81_ma107_nor100},
\ref{Barish_77_ma95_nor100}, \ref{Miller_83_ma100_nor97},
and \ref{Kit_83_ma105_nor95}.  In these figures, our curves are normalized 
such that they match the previous curves at one point in the low $Q^2$ region,
so that we can compare the $Q^2$-dependence of the spectra.
The y-axis is weighted events/bin (corrected for efficiencies).
We reproduce the shape ($Q^2$-dependence) of the calculations of
Baker {\em et al.}~\cite{Baker_81},
Barish {\em et al.}~\cite{Barish_77}, 
and Kitagaki {\em et al.}~\cite{Kitagaki_83}, but not
Miller {\em et al.}~\cite{Miller_82}. 
As Miller {\em et al.} gives an updated result of 
Barish {\em et al.}, they should be using the same calculation.
Therefore, we do not understand the origin of the curve shown
in Miller {\em et al.} 

Having reproduced the calculations under the same assumptions as in the
original extractions, we perform our own extraction of $M_A$, using our
calculation of the cross sections, but using the same input ($g_A$ and form
factors) as assumed in the original extractions.
Due to inefficiencies in reconstruction for very low $Q^2$ events,
they do not use the first bin for fitting. Hence, we do not 
use the first bin for fitting or normalization either.
We perform a binned maximum likelihood using a formula from 
the Particle Data Group~\cite{pdg}. 
The experiments performed an unbinned likelihood fit,
which we cannot reproduce since we do not have the individual events.
Table~\ref{MA_values} gives the results of these fits. 
Using the same assumptions, we reproduce the fitted value of $M_A$
from  Baker {\em et al.}, while disagreeing somewhat with the values from 
Barish {\em et al.}, Miller {\em et al.}, and Kitagaki {\em et al.}
%Barish {\em et al.} and  Miller {\em et al.} quote
%a `$Q^2$ shape' $M_A$ and a flux-independent  $M_A$.
%Their final quoted value (the values on their plot) 
%is the flux-independent $M_A$. However, both  methods use unbinned 
%likelihood. Perhaps their unbinned likelihood fit as 
%opposed to our binned likelihood fit creates the difference.
The difference may come from the fact that we are forced to use a binned
likelihood fit, rather than being able to reproduce their unbinned fit.

%figure13
\begin{figure}
\begin{center}
\epsfxsize=2.91in
\mbox{\epsffile[95 260 540 580]{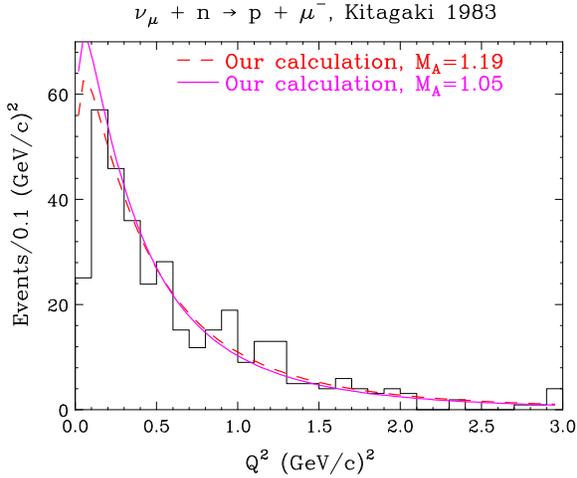}}
\end{center}
\caption{ $Q^2$ distribution from Kitagaki {\em et al.}~\cite{Kitagaki_83}.
The dash curve is our calculation using our fit value of  $M_A$=1.19 GeV. 
The solid curve is our calculation using their fit value of $M_A$=1.05 GeV. }
\label{Kit_83_o0oo_105_119_100}
\end{figure}
%figure 14
\begin{figure}
\begin{center}
\epsfxsize=2.91in
\mbox{\epsffile[95 260 540 580]{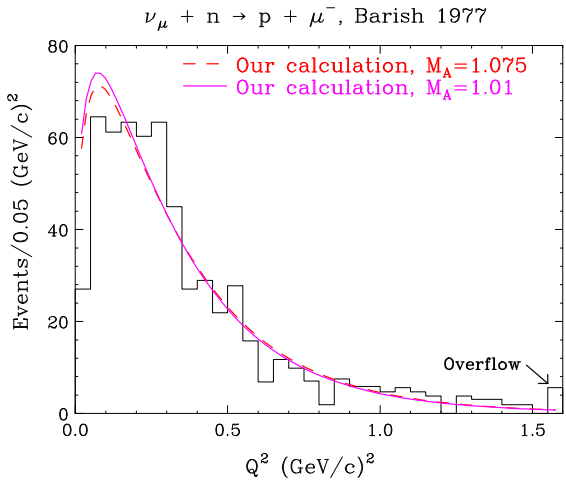}}
\end{center}
\caption{ $Q^2$ distribution from Barish {\em et al.}~\cite{Barish_77}.
The dash curve is our calculation using our fit value of  $M_A$=1.075 GeV. 
The solid curve is our calculation using their fit value of $M_A$=1.01 GeV.  }
\label{Barish_77_o0oo_107_101_100}
\end{figure}

\begin{figure}
\begin{center}
\epsfxsize=2.91in
\mbox{\epsffile[95 260 540 580]{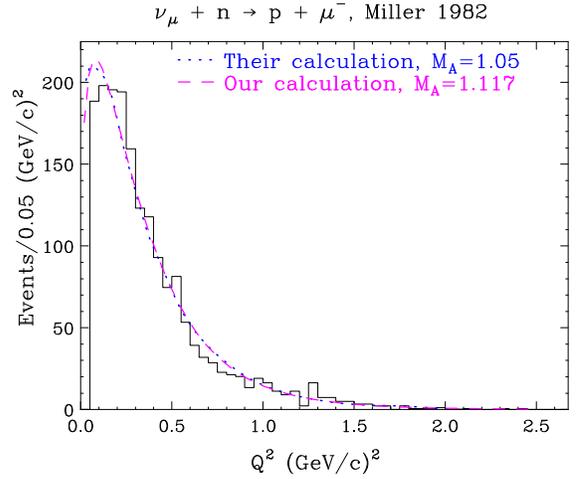}}
\end{center}
\caption{ $Q^2$ distribution from Miller {\em et al.}~\cite{Miller_82}.
The dotted curve is their calculation using 
their fit value of $M_A$=1.05 GeV. The
dash curve is our calculation using our fit value of  $M_A$=1.117 GeV. }
\label{Miller_82_o0oo_105_119_100}
\end{figure}

Figures~\ref{Kit_83_o0oo_105_119_100} and~\ref{Barish_77_o0oo_107_101_100}
show the difference between using their value of 
$M_A$ and our value of $M_A$ for Kitagaki {\em et al.} and  
Barish {\em et al.}  The plot for  Kitagaki {\em et al.} 
appears to show that our value of $M_A$=1.19 GeV
is a better fit than their value of $M_A$.
For Barish {\em et al.} its not clear which is a better fit. 
As previously stated, our $Q^2$  distribution is slightly
different than that of Miller {\em et al.} for the same $M_A$.
Figure~\ref{Miller_82_o0oo_105_119_100} shows their 
calculation for their best fit $M_A$ versus our calculation
using our best fit $M_A$. The two shapes agree very well.
Hence, we are able to reproduce the  best fit shape of Miller {\em et al.} 
fit, but not with their value of $M_A$.

%figure15

%Table 3
\begin{table*} 
\begin{center}
\begin{tabular}{|l|c|c|c|c|c|c|}
\noalign{\vspace{-8pt}} \hline
            & $M_A$             & updated $M_A$ & updated $M_A$   & $\Delta M_A$ & $\Delta M_A$ \\
            & (published)       & old params.   & new params.     & new--old    & BBA-2003--Dipole \\ \hline
 Baker 1981 \cite{Baker_81}       & 1.07 $\pm$ 0.06      &1.079 $\pm$ 0.056 & 1.055 $\pm$ 0.055 & $-0.024$ & $-0.049$ \\ 
 Barish 1977 \cite{Barish_77}     & 1.01 $\pm$ 0.09      &1.075 $\pm$ 0.10 & 1.049 $\pm$ 0.099 & $-0.026$  & $-0.046$ \\  
 Miller 1982 \cite{Miller_82}     & 1.05 $\pm$ 0.05      &1.117 $\pm$ 0.055& 1.090 $\pm$ 0.055 & $-0.027$  & $-0.046$ \\
 Kitagaki 1983 \cite{Kitagaki_83} & 1.05$_{-0.16}^{+0.12}$&1.194$_{-0.11}^{+0.10}$&1.175$_{-0.11}^{+0.10}$ & $-0.019$  & $-0.050$ \\ \hline
\end{tabular}
\end{center}
\caption{Published and updated extractions of $M_A$ (GeV) from deuterium experiments. 
The first value of $M_A$ is the values extracted in the original publications.
For Barish and Miller, we give their `shape fit' value,
since this value most closely reflects how we can calculate their $M_A$.
The second value of $M_A$ is from the analysis presented here,
using the same input parameters (form factors and $g_A$) as in the publications,
while the third uses the updated parameters from tables~\ref{JRA_coef} and
\ref{parameters}.  The last two columns show the change in $M_A$ between the new 
and old input parameters, and the change when comparing the BBA-2003 and Dipole Form
Factors (with $g_A$ fixed).}
%\caption{Fit values and shifted values of $M_A$ (GeV) from deuterium experiments. 
%Column 2 gives the fit values of $M_A$ from
%their papers. For Barish and Miller, we give their `shape fit' value,
%since this value most closely reflects how we can calculate their $M_A$.
%Column 3 gives our fit
%value of $M_A$ using their assumptions. Column 4 gives our fit value of 
%$M_A$ with the BBA-2003 Form Factors.
%Column 5 gives $\delta M_A$    
%between our assumptions minus the experiments 
%assumptions. Column 6 gives $\delta M_A$
%between using BBA-2003 Form Factors and  
%Dipole Form Factors, with $g_A$ constant.}
\label{MA_values}
\end{table*}

Finally, we extract $M_A$ for each of these experiments using
our calculations with the updated BBA-2003 Form Factors and $g_A$ value. 
Comparing this to our extraction with their input parameters, we obtain
change in $M_A$ due to using updated values for $g_A$ and the form factors.
Because we sometimes obtain slightly different values of $M_A$, even with the
same input parameters, we take the difference between our extraction with old
and new form factors as the modification that should be applied to the previous
extractions, which were able to do a more detailed comparison to their data.
Table~\ref{MA_values} gives the results of these fits, which indicate
that we should shift
the value of $M_A$ determined from deuterium by $-0.025$ GeV
from the value quoted by these experiments. We also
show that a shift in $M_A$ of $-0.050$ GeV is required in going from the Dipole
Form Factors to BBA-2003 Form Factors, keeping $g_A$ constant. If only cross 
section data were used for the form factor fits (Table~\ref{parameters},
rows 4 and 5), the  value of $M_A$ would go up by 0.002 GeV, a small effect.

%figure16
\begin{figure}
\begin{center}
\epsfxsize=2.91in
\mbox{\epsffile[95 260 540 580]{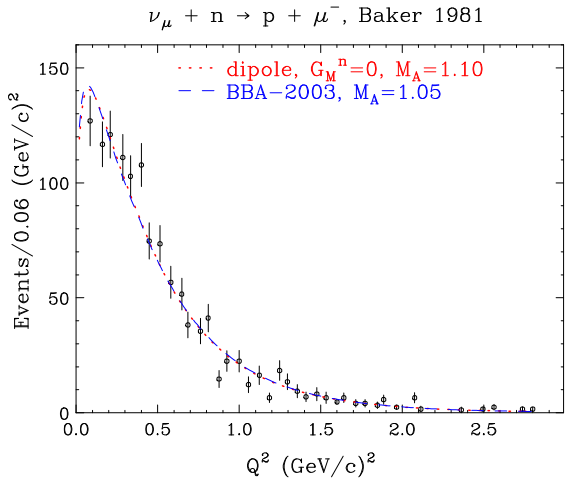}}
\end{center}
\caption{ A comparison of the $Q^2$ distribution using 2 different
sets of form factors. The data are from Baker {\em et al.}~\cite{Baker_81}.
The dotted curve uses Dipole Form Factors with 
$M_A$=1.10 GeV. The dashed curve uses BBA-2003 Form Factors
with $M_A$=1.05 GeV. }
\label{Baker_d0dd_110_JhaKJhaJ_105}
\end{figure}

Figure~\ref{Baker_d0dd_110_JhaKJhaJ_105} shows the $Q^2$ distribution
for Dipole Form Factors and $M_A$=1.10 GeV
with the distribution for BBA-2003 Form Factors
and $M_A$=1.050 GeV.  When we modify the electromagnetic form factors,
the modification in $M_A$ not only reproduces the original yield,
it also reproduces the $Q^2$ distribution.
Because there is no modification of the $Q^2$ dependence when strength
is shifted between the electromagnetic and axial form factors, we conclude
that the use of Dipole Form Factors will lead to an error in $M_A$ of 0.050
GeV, independent of the details of the experiment.

%figure17
\begin{figure*}
\begin{center}
\epsfxsize=6.51in
\mbox{\epsffile[0 0 567 240]{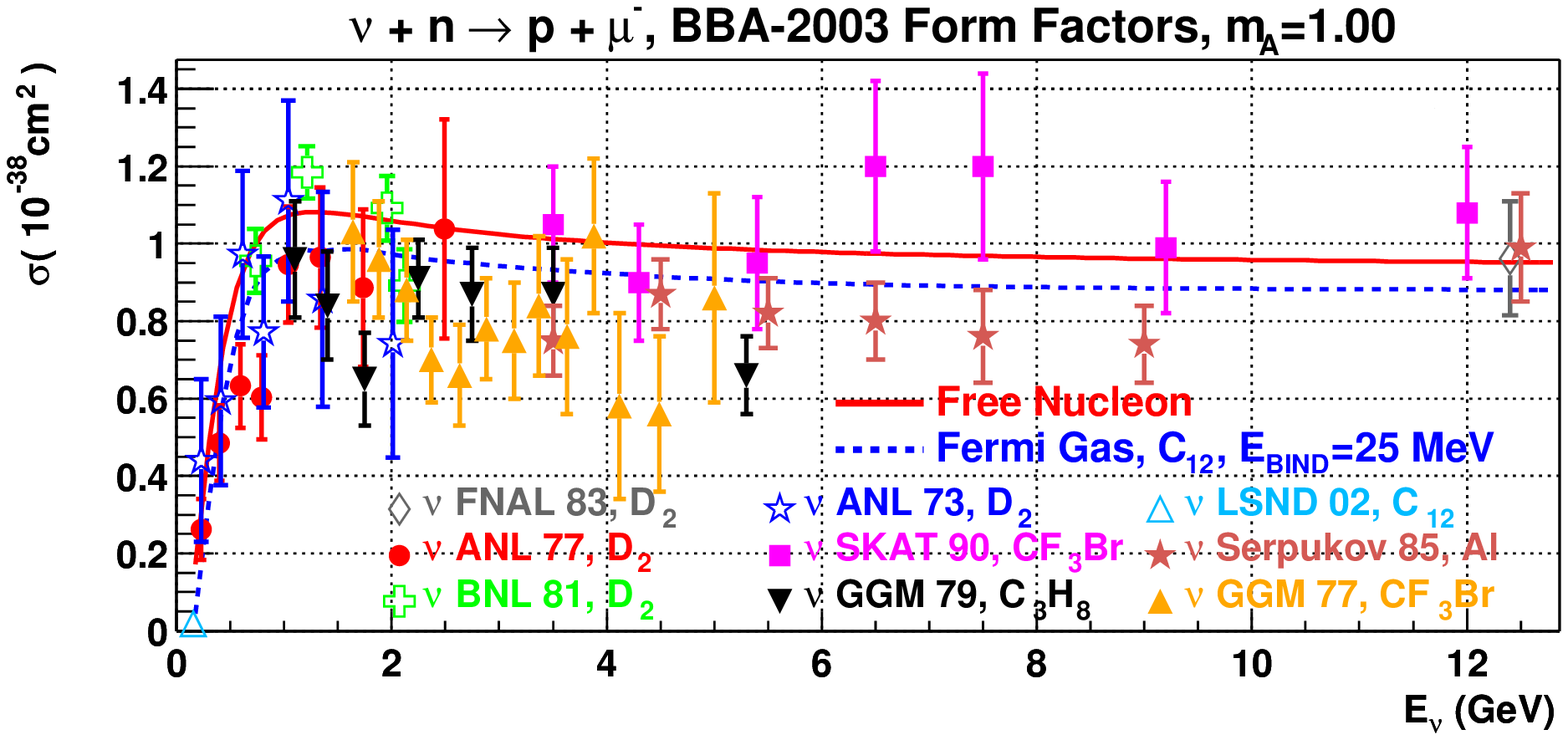}}
\epsfxsize=6.51in
\mbox{\epsffile[0 30 567 260]{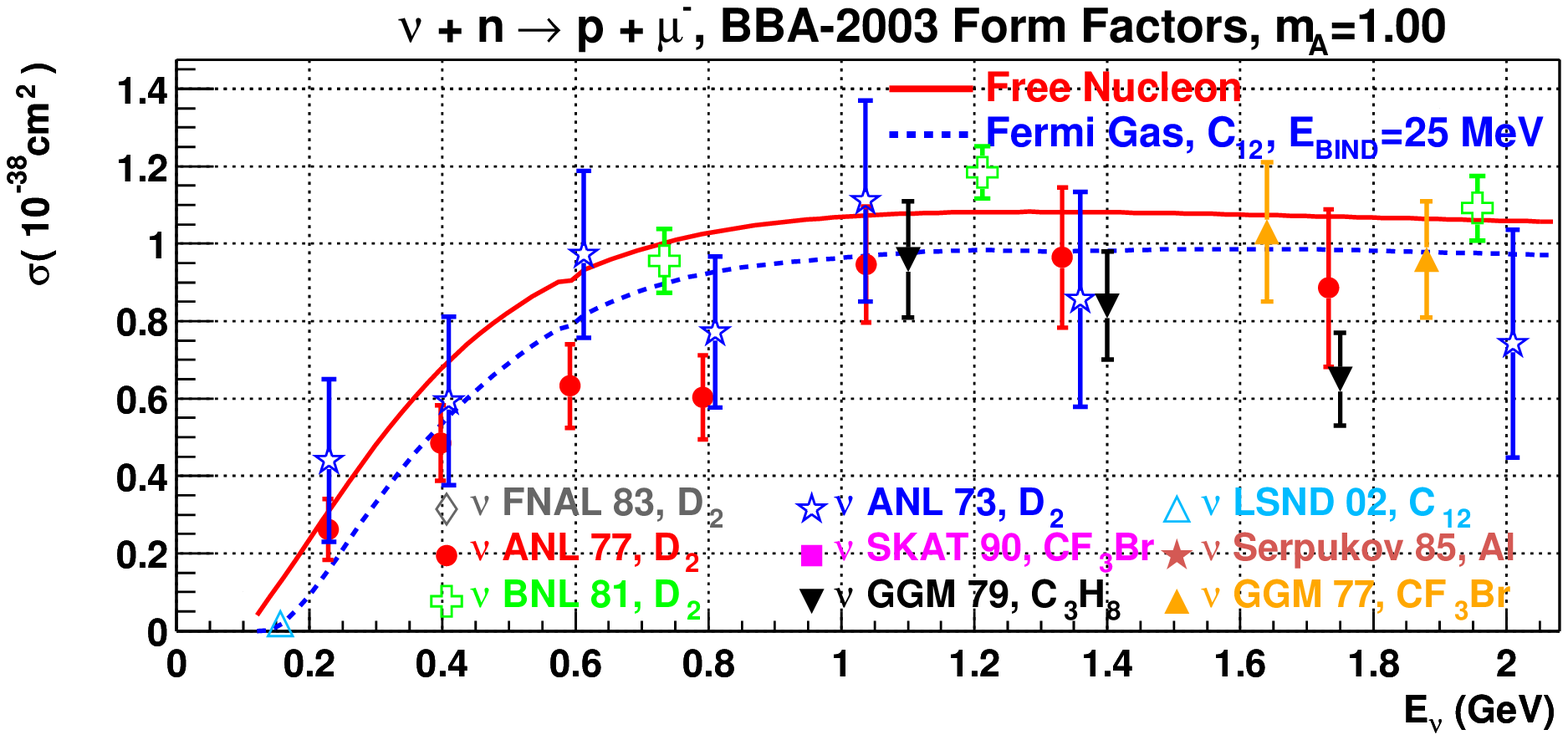}}
\end{center}
\caption{The QE neutrino cross section along 
with data from various experiments. The calculation uses 
$M_A$=1.00 GeV, $g_A$=$-1.267$, $M_V^2$=0.71 GeV$^2$ and 
BBA-2003 Form Factors.
The solid curve uses no nuclear correction, 
while the dotted curve~\cite{Zeller_03} uses
a Fermi gas model for carbon with a 25 MeV binding energy 
and 220 Fermi momentum.
 The lower plot is identical to the 
upper plot with the $E_{\nu}$ axis limit changed to 2~GeV.
The data shown are from 
FNAL 1983~\cite{Kitagaki_83},
ANL 1977~\cite{Barish_77},
BNL 1981~\cite{Baker_81},
ANL 1973~\cite{Mann_73},
SKAT 1990~\cite{Brunner_90},
GGM 1979~\cite{Pohl_79},
LSND 2002~\cite{Auerbach_02},
Serpukov 1985~\cite{Belikov_85},
and GGM 1977~\cite{Bonetti_77}.}
\label{elas_JhaKJhaJ_nu}
\end{figure*}

%figure18
\begin{figure*}
\begin{center}
\epsfxsize=6.51in
\mbox{\epsffile[0 35 567 255]{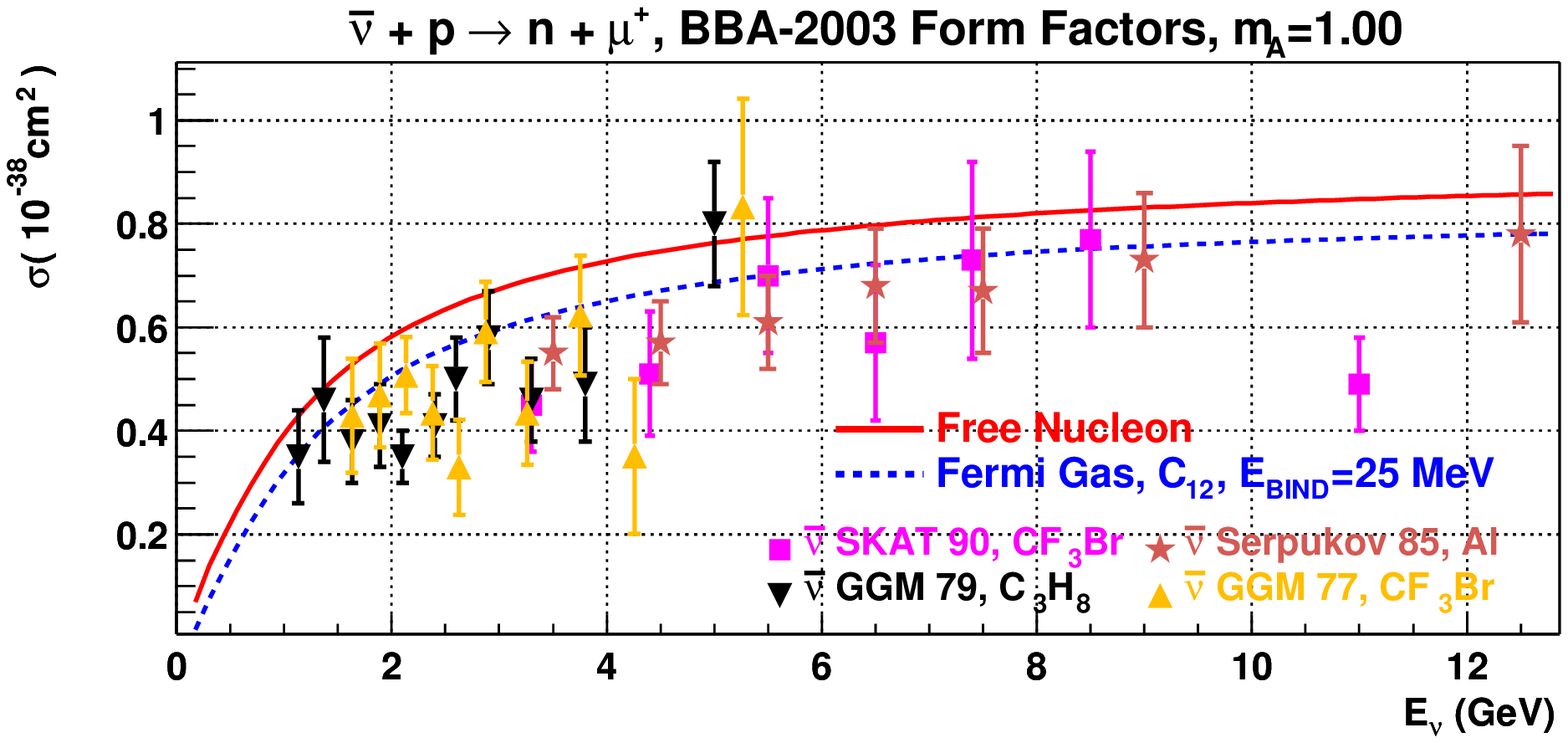}}
\end{center}
\caption{  The QE antineutrino cross section along 
with data from various experiments.
 The calculation uses 
$M_A$=1.00 GeV, $g_A$=$-1.267$, $M_V^2$=0.71 GeV$^2$ and BBA-2003 Form Factors.
The solid curve uses no nuclear correction, 
while the dotted curve~\cite{Zeller_03} uses
a Fermi gas model for carbon with a 25 MeV 
binding energy and 220 MeV Fermi momentum.
The data shown are from  
SKAT 1990~\cite{Brunner_90}, 
GGM 1979~\cite{Armenise_79},
Serpukov 1985~\cite{Belikov_85},
and GGM 1977~\cite{Bonetti_77}.}
\label{elas_JhaKJhaJ_nub}
\end{figure*}

Figures~\ref{elas_JhaKJhaJ_nu} and~\ref{elas_JhaKJhaJ_nub} show the QE cross section for 
$\nu$ and $\overline{\nu}$ using our most up to date assumptions.
The normalization uncertainty in the data is approximately 10\%.
We have used BBA-2003 Form Factors, and have scaled 
down $M_A$ from the old best fit of $M_A$=1.026 $\pm$ 0.021 GeV to $M_A$=1.00 GeV
(which would have been obtained with the best vector form factors known today). 
The solid curve uses no nuclear correction, while the dotted curve~\cite{Zeller_03} uses
a NUANCE~\cite{Casper_02} calculation of a  Smith and Moniz~\cite{Smith_72} based Fermi gas model
for carbon. This nuclear model includes Pauli blocking and Fermi
motion, but not final state interactions. 
The Fermi gas model was run with a 25 MeV binding energy and
220 MeV Fermi momentum.
The updated form factors improve the agreement with neutrino QE cross section data 
and give a reasonable description of the cross sections from deuterium.  However,
the data from heavy targets, including all of the anti-neutrino data, are
systematically below the calculation, even with the Fermi gas nuclear corrections.  
The only experiment on heavy nuclei which agrees with the calculation
is SKAT~\cite{Brunner_90} with neutrinos.
The Fermi gas nuclear correction may not be sufficient.
Tsushima  {\em et al.}~\cite{Tsushima_03} studied the effect of 
nuclear binding on the nucleon form factors. They stated that 
modifications in bound nucleon 
form factors reduce the cross section relative to 
free nucleon form factors by 8\%.
We plan to study the nuclear corrections, adopting models
which have been used in precision electron scattering measurements
from nuclei at SLAC and JLab.  In addition, we will
be updating the extraction of $M_A$ from other experiments,
using the updated versions of the input parameters and electromagnetic form factors.

%figure19
\begin{figure}
\begin{center}
\epsfxsize=2.91in
\mbox{\epsffile[7 75 550 340]{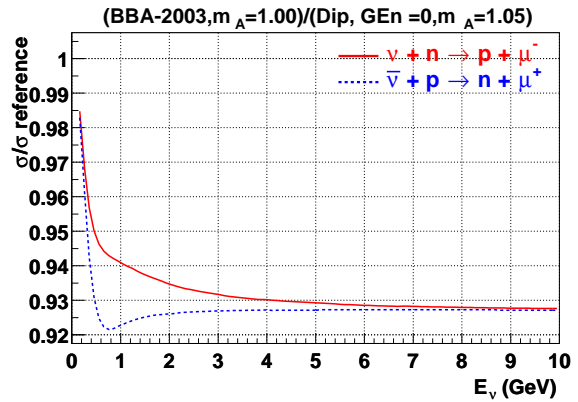}}
\end{center}
\caption{ Ratio of cross section versus energy using  BBA-2003 
Form Factors with $M_A$=1.00 GeV versus Dipole Form Factors with $M_A$=1.05 GeV. } 
\label{r_JhaKJhaJ_ma100_D0DD_ma105}
\end{figure}
%figure20
\begin{figure}
\begin{center}
\epsfxsize=2.91in
\mbox{\epsffile[7 75 550 360]{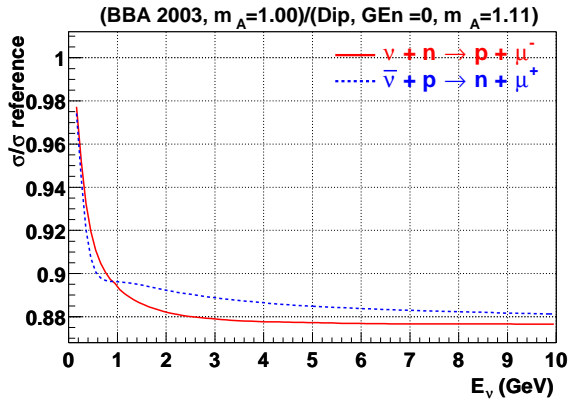}}
\end{center}
\caption{ Ratio of cross section versus energy using  BBA-2003 
Form Factors with $M_A$=1.00 GeV versus Dipole Form Factors with $M_A$=1.11 GeV.}
\label{r_JhaKJhaJ_ma100_D0DD_ma111}
\end{figure}

Both the overall cross section and the $Q^2$-dependence depend on the form
factors chosen.  However, even for input form factors that yield identical
$Q^2$-dependences, the cross section can differ significantly.
Figure~\ref{r_JhaKJhaJ_ma100_D0DD_ma105}
compares the cross section for two sets of form factors with 
identical $Q^2$ distribution (shown in Fig.~\ref{Baker_d0dd_110_JhaKJhaJ_105}).
The figure shows that
a cross section error as large as 7.5\% can occur from using this
combination of incorrect form factors, even though they match
the `$Q^2$ shape' as well as the correct form factors do.  The effects can
be even larger than in this example, especially if the $Q^2$-dependence
is not required to be identical.
As shown by Y. Itow's at NUINT02 talk on the results from 
K2K~\cite{nuint02}, the initial K2K analysis (on a water target)
assumed Dipole Form Factors.  They obtained a value of $M_A$=1.11 GeV
and saw a low $Q^2$ suppression with respect to the predicted distribution
with Dipole Form Factors and $M_A = 1.023$.  Figure~\ref{r_JhaKJhaJ_ma100_D0DD_ma111}
shows the ratio vs. energy of cross sections predicted using 
BBA-2003 Form Factors with $M_A$=1.00 GeV 
to the prediction with Dipole Form Factors with $M_A$=1.11 GeV.
The figure shows differences as large as 12\%. This will clearly be
important for future neutrino experiments at Fermilab and elsewhere:
Even if the form factors which are used in the model
are adjusted to describe the $Q^2$ distribution measured
in a near detector neutrino experiment, one
would be predicting an incorrect energy dependence of
the QE cross section. Therefore, using the correct
combination of form factors is
important for determining the neutrino cross section
and its energy dependence.  An accurate 
measurement of QE scattering cross sections  requires
an accurate measurement
of both the normalized cross section versus energy
as well as the shape of the $Q^2$ distribution.

\section{CONCLUSION}
  We have made an updated extraction of nucleon electromagnetic form 
factors electron scattering, and have shown these have as much
as a 6\% effect for neutrino-nucleon QE scattering when
compared with the standard dipole approximation.  Inclusion of
the new form factors yields to a reduction in the value of $M_A$ 
extracted from neutrino scattering of 0.025 GeV.
The agreement between the calculated 
neutrino and antineutrino free nucleon
cross section and data is improved using the updated form factors,
but is not spectacular, especially for data taken on heavy
targets.
We have shown that for a fixed 
$Q^2$ dependence, the 
neutrino cross section and its energy dependence
can be affected by more than 10\%
if the input shapes of the electromagnetic form 
factors used are not correct.  Hence, a complete 
understanding of QE scattering scattering requires
an accurate measurement  
of both the normalized cross section versus energy
as well as the shape of the $Q^2$ distribution.  In addition,
nuclear effects such as Pauli blocking and modification
of nucleon form factors in bound nuclei need to be
included. These effects are currently under investigation. For example, the
simple Fermi gas model can be modified to include
a high momentum tail~\cite{Bodek_81} or more sophisticated spectral functions.

\section{ACKNOWLEDGMENTS}
This work is supported in part by the U. S. Department of Energy, Nuclear
Physics Division, under contract W-31-109-ENG-38 (Argonne)
and High Energy Physics Division under
grant DE-FG02-91ER40685 (Rochester).

\end{document}